\documentstyle[12pt]{article}
\begin{document}
\begin{titlepage}
\title{$GL_{q}(N)$-Covariant Quantum Algebras \\
and Covariant Differential Calculus\thanks{Preprint Dubna, JINR
E2-92-477 (1992)} }
\author{\em  A.P. Isaev\thanks{e-mail address: isaevap@theor.jinrc.dubna.su}
 and P.N.Pyatov\thanks{e-mail address: pyatov@theor.jinrc.dubna.su} \\
\rm Laboratory of Theoretical Physics, \\
JINR, Dubna, SU-101  000 Moscow, Russia}
\date{}
\maketitle
\begin{abstract}
We consider $GL_{q}(N)$-covariant quantum algebras with generators
satisfying quadratic polynomial relations. We show that,
up to some inessential arbitrariness, there are
only two kinds of such quantum algebras, namely, the algebras
with $q$-deformed commutation and $q$-deformed anticommutation
relations. The connection with the bicovariant differential
calculus on the linear quantum groups is disscussed.
\end{abstract}
\end{titlepage}
\newpage
\section{Introduction}
\setcounter{equation}0

   Noncommutative geometry \cite{Conne}
has awakened increasing interest and
has started to play a very significant role
in mathematical physics for last few years. The
attractive field of investigations here is the theory of
quantum groups \cite{Drin}-\cite{FRT}
and especially several differential geometric aspects of this theory
such as differential calculus
on the quantum groups. A bicovariant version of this calculus
has been formulated in
the general form by S.L.Woronowicz \cite{Woron2}.
Then, an intimate relation of the Woronowicz's
bicovariant calculus with R-matrix formalism for the quantum groups
\cite{FRT} has
been established in Refs. \cite{Jurco,Zumino}. Quite
recently, a systematic realization of
the bicovariant differential calculus in the
framework of the R-matrix approach
has been achieved in \cite{Faddeev}.
These results give us the promising possibility to
use the quantum groups as generalizations of the classical
symmetry groups in various physical models.

In this paper we realize the ideas of  Refs.
\cite{Woron2}-\cite{Faddeev}
and derive explicit formulas for $GL_{q}(N)$ ($SL_{q}(N)$)-bicovariant
differential calculus by means of
considering quantum algebras which are covariant under the
coaction of $Fun(GL_{q}(N))$\footnote{Further we use the short
notation $GL_{q}(N)$ instead of $Fun(GL_{q}(N))$.}.
The starting point of our considerations
is the observation that
right(left)-invariant vector fields $E^{i}_{j}$
and differential
1-forms $\Omega^{i}_{j}$
\footnote{Here the elements $E^{i}_{j}$ or $\Omega^{i}_{j}$
($i,j \in 1, \ldots ,N$) form the basis in
the space of right(left)-invariant vector fields or 1-forms,
respectively.}
on $GL_{q}(N)$ can be treated as elements of the adjoint
$GL_{q}(N)$-comodules or, in other words,
they realize the adjoint representations
of $GL_{q}(N)$ in the sense of Ref.\cite{FRT}.
Then, we consider the general associative algebras with unity
whose generating elements $A^{i}_{j}$
(the unified notation for $E^{i}_{j}$ or $\Omega^{i}_{j}$)
are constrained by
certain quadratic polynomial relations.
We require these relations to be covariant
under the transformations of $A^{i}_{j}$ as
the adjoint $GL_{q}(N)$-comodule
($T^{i}_{j} \in GL_{q}(N)$)
\begin{equation}
A^{i}_{j} \rightarrow T^{i}_{i^{'}}
S(T)^{j^{'}}_{j} \otimes A^{i^{'}}_{j^{'}}
\equiv (TAT^{-1})^{i}_{j} \; .
\label{a1}
\end{equation}
In the last part of
(\ref{a1}) the short notation is introduced
to be used below. Besides, we demand that the
quadratic polynomial relations for
$A^{i}_{j}$ allow us to make the lexicographic
ordering for any monomial of the type
$A^{i_{1}}_{j_{1}}A^{i_{2}}_{j_{2}} \cdots A^{i_{n}}_{j_{n}}$.
Later on we
refer to the algebras with such features
as the $GL_{q}(N)$-covariant
quantum algebras.

The
quadratic polynomial
relations for $GL_{q}(N)$-covariant
quantum algebras
can be written in the following general form
\begin{equation}
\langle \alpha |^{jl}_{ik} \rangle
A^{i}_{j}A^{k}_{l} =
\langle \alpha |^{m}_{n} \rangle
A^{n}_{m} + C(\alpha ) ,
\label{a2}
\end{equation}
where the index $\alpha$  enumerates  different
relations and
the coefficients
$\langle \alpha |^{jl}_{ik} \rangle$ ,
$\langle \alpha |^{m}_{n} \rangle $ and
$C(\alpha)$ are
functions of the deformation parameter $q$.
On the condition that Eqs.(\ref{a2}) are
covariant under transformations (\ref{a1}) we obtain
that parameters
$\langle \alpha |^{jl}_{ik} \rangle $
are  $q$-analogs of the Clebsch-Gordon coefficients coupling
two adjoint $GL_{q}(N)$ representations
into the irreducible representations (irreps). Parameters
$\langle \alpha |^{m}_{n} \rangle $
can be considered as harmonics which are not equal to zero only if
$\langle \alpha |^{jl}_{ik} \rangle $
couple $A \otimes A$ into the adjoint $GL_{q}(N)$-comodule again, while
$C(\alpha ) \neq 0$
only if combination
$\langle \alpha |^{jl}_{ik} \rangle
A^{i}_{j}A^{k}_{l}$
is expressed in terms of Casimir operators.
Here we use the idea that
arbitrary monomials
$A^{i_{1}}_{j_{1}}A^{i_{2}}_{j_{2}} \cdots A^{i_{n}}_{j_{n}}$
(transformed in accordance with (\ref{a1}))
can be considered as components of
$GL_{q}(N)$-tensor operators.
Some papers have already
appeared in which tensor
operators for quantum
groups are discussed in another context \cite{BiedRitt}.

We find that, up to some arbitrariness discussed in Sect.3,
there are only two kinds
of $GL_{q}(N)$-covariant quantum
algebras. For the first one the left-hand
side of Eq.(\ref{a2}) is the $q$-deformed
commutator while for the second
one it has the form of $q$-deformed anticommutator.
It is natural to call the
algebras of the first and  second
kind as "bosonic" and "fermionic" $GL_{q}(N)$-covariant quantum
algebras and relate them with the algebras of
right(left)-invariant vector fields and 1-forms on
$GL_{q}(N)$, respectively.
As we shall see, these conjectures are justified by
some  explicit construction for the differential calculus on
$GL_{q}(N)$ and are in agreement with the results
obtained in Refs. \cite{Woron2}-\cite{Faddeev}.

\section{R-matrix formulation of $GL_{q}(N)$ and
$GL_{q}(N)$- covariant commutator and anticommutator.}
\setcounter{equation}0

This section is a review of some facts about quantum groups
needed in the consideration below. We follow the approach by Faddeev,
Reshetikhin and Takhtajan \cite{FRT}. The generators
of the quantum group $GL_{q}(N)$
can be defined as elements of N by N matrix
$T^{i}_{j}$ $(T \in Mat(N,C))$
with commutation relations
\begin{equation}
R_{12}T_{1}T_{2}=T_{2}T_{1}R_{12} \; .
\label{b1}
\end{equation}
Here and henceforth we use the notation
of Ref.\cite{FRT}. The R-matrix for
$GL_{q}(N)$ looks like \cite{Jimbo}
\begin{equation}
R_{12} =
 R^{i_{1} i_{2} }_{j_{1} j_{2} }
= \delta ^{i_{1}}_{j_{1}} \delta ^{i_{2}}_{j_{2}}
\left( 1+(q-1) \delta ^{i_{1} i_{2} } \right) + (q-q^{-1})
\delta ^{i_{1}}_{j_{2}} \, \delta
^{i_{2}}_{j_{1}} \,\theta (i_{1} - i_{2} )
\label{b2}
\end{equation}
where
$
\theta (i-j) = \left \{
\begin{array}{ll}
1, & i>j \\ 0, & i \leq j
\end{array}
\right.
$.
The associativity conditions for the
relations (\ref{b1}) yield the Yang-Baxter
equation for the R-matrix
\begin{equation}
R_{12}R_{13}R_{23} = R_{23}R_{13}R_{12}
\Longleftrightarrow
R_{12}R^{-1}_{31}R^{-1}_{32}= R^{-1}_{32}R^{-1}_{31}R_{12}
\label{b3}
\end{equation}
Comparing (\ref{b1}) and (\ref{b3}) we see that
possible matrix realizations of the operators $T^{i}_{j}$ are
\begin{equation}
\label{b4}
(T^{i}_{j})^{k}_{l} = R^{ik}_{jl} \;, \;\;
(T^{i}_{j})^{k}_{l} = (R^{-1})^{ki}_{lj}\;.
\end{equation}

The R-matrix (\ref{b2}) obeys the Hecke relation which
can be rewritten as
\begin{equation}
R_{21} = (q-q^{-1})P_{12} + R^{-1}_{12} \;.
\label{b9a}
\end{equation}
where
$(P_{12})^{i_{1}i_{2}}_{j_{1}j_{2}} =
\delta^{i_{1}}_{j_{2}} \delta^{i_{2}}_{j_{1}} $
is the permutation matrix.
According to Eq. (\ref{b9a}) one can define two
projectors
\begin{equation}
P^{\pm }_{12} =
\frac{P_{12}}{q+q^{-1}}( R_{12} \pm
q^{ \pm 1}P_{12}) \; ,
\label{b10}
\end{equation}
which are quantum analogs of the
symmetrizer $\frac{1}{2} (I+P_{12})$
and antisymmetrizer $\frac{1}{2}
(I-P_{12})$. Here
 $(I)^{i_{1} i_{2}}_{j_{1} j_{2}}
= \delta^{i_{1}}_{j_{1}}  \delta^{i_{2}}_{j_{2}} $
is the identity matrix.
As it has been shown in Refs.\cite{LusRos},
if $q$ is not a root of unity,
the representation theory for $GL_{q}(N)$ can be
constructed in the same way as for $GL(N)$. Indeed,
with the help of the projectors (\ref{b10}) one can
construct $q$-analogs of the
Young operators of symmetrization \cite{Jimbo,Weyl}
and thus realize the
program of extracting irreducible $GL_{q}(N)$- and
$SL_{q}(N)$-comodules from the direct product of the fundamental
comodules.

Let us demonstrate this by decomposing
the direct product of two adjoint comodules.
The method coincides in principle with
the well known prescription for decomposing the
direct product of two mesonic
representations considered in the framework of the $SU(N)$-quark
models of strong interactions
(see e.g. remarkable reviews \cite{Smorod,Bog}).
First, we note that the tensor $A^{i}_{j}$
has $N^{2}$
components and it is possible to
decompose it into
the scalar $Tr_{q}(A)$ and
the $q$-traceless tensor
$\tilde{A} ^{i}_{j}$
with $(N^{2}-1)$
independent components
\begin{equation}
\tilde{A}^{i}_{j} = A^{i}_{j} -
\delta^{i}_{j} Tr_{q}(A)/
(\sum_{i=1}^{N} q^{2i}) \; .
\label{b11}
\end{equation}
Here we have introduced the $q$-deformed
trace \cite{FRT,Zumino,Faddeev,Isaev}
\begin{equation}
Tr_{q}A \equiv Tr(DA) \equiv \sum_{i=1}^{N} q^{2i} A^{i}_{i}
\label{b12}
\end{equation}
 satisfying the following
invariance property ($T^{i}_{j} \in GL_{q}(N)$)
\begin{equation}
\label{b13}
Tr_{q}(A) \rightarrow Tr_{q}(TAT^{-1}) = Tr_{q}A \;
\end{equation}
which is true for any matrix representation of $T^{i}_{j}$,
in particular, for (\ref{b4}).
Using the construction of the $q$-trace one
can reproduce the $GL_{q}(N)$-invariants as
\begin{equation}
C_{n} = Tr_{q}(A^{n}) \;, \;\;\;\;\;\;\;\;
n \geq 1 \; .
\label{b14}
\end{equation}
Now we introduce the basic covariant bilinear combinations
$P_{12}A_{1}R_{21}A_{2}$ of
tensors $A^{i}_{j}$ with the transformation rule
\begin{equation}
P_{12}A_{1}R_{21}A_{2} \rightarrow
T_{2}T_{1}(P_{12}A_{1}R_{21}A_{2})T^{-1}_{1}T^{-1}_{2} \; .
\label{b15}
\end{equation}
Using projectors (\ref{b10}) it is
possible to decompose tensor (\ref{b15}) into
the four independently transformed tensors
\begin{equation}
X^{\pm \pm}_{q}  =
P^{\pm }_{21} ( P_{12}A_{1}R_{21}A_{2} ) P^{\pm }_{21} \; ,
\;\; X^{\pm \mp}_{q}  =
P^{\pm }_{21} (P_{12}A_{1}R_{21}A_{2} ) P^{\mp }_{21} \; .
\label{b16}
\end{equation}
The dimensions of these $GL_{q}(N)$-comodules are
$\frac{N^{2}(N+1)^{2} }{2}$ (for $X^{++}_{q}$),
$\frac{N^{2}(N-1)^{2} }{2}$ (for $X^{--}_{q}$)
and $\frac{N^{2}(N^{2} - 1) }{2}$ (for $X^{\pm \mp}_{q}$). Their
undeformed ($q$=1) analogs are nothing but
\begin{equation}
X^{\pm \pm}  =
\frac{1}{4} (P_{12} \pm I) [ A_{1}, A_{2} ]_{+} \; , \;\;
X^{\pm \mp}  =
\frac{1}{4} (P_{12} \pm I) [ A_{1}, A_{2} ]_{-} \; .
\label{b17}
\end{equation}
As it is seen from (\ref{b17}),
$X^{\pm \pm}$ are expressed in terms of the anticommutators,
while $X^{\pm \mp}$ yield the combinations of the commutators.
On the other hand, one can express the
commutator and anticommutator as linear
combinations of $X^{\pm\mp}$ and $X^{\pm\pm}$
as given below
\begin{equation}
X^{+-} - X^{-+}  =  \frac{1}{2} [ A_{1}, A_{2} ]_{-} \; ,
\;\; X^{++} - X^{--}  =  \frac{1}{2} [ A_{1}, A_{2} ]_{+} \; .
\label{b20}
\end{equation}
It is worth noting here that linear combinations of
$X^{++}$ with $X^{--}$ or $X^{+-}$ with $X^{-+}$ are
the only two possibilities to obtain for any pair of generators
$A^{i}_{j}$, $A^{k}_{l}$ the bilinear expressions
of the type $[ A^{i}_{j}, \; A^{k}_{l} ]_{\alpha} =
A^{i}_{j} A^{k}_{l} - \alpha A^{k}_{l} A^{i}_{j} \; (\alpha \neq 0)$
which can be used as the left-hand side of (\ref{a2}) ($q=1$). Only
such quadratic polynomial relations allow us
to reorder any monomial $A^{i}_{j} \cdots A^{k}_{l}$ in an
appropriate way (see Sect.1).
Indeed, combining, for example,
$X^{++}$ with $X^{+-}$ or $X^{--}$ with  $X^{-+}$ we
are unable to commute
$A^{i}_{j}$ and $A^{k}_{l}$ when $j=l$, while the
combinations of $X^{++}$ and $X^{-+}$
or $X^{--}$ and $X^{+-}$ are
unsatisfactory for reordering the pairs $A^{i}_{j}$,
$A^{k}_{l}$ when $k=i$. So, it seems reasonable
to use only $X^{++}_{q}$
together with $X^{--}_{q}$ or
$X^{+-}_{q}$ together with $X^{-+}_{q}$
in defining relations
(\ref{a2}) in order to solve the ordering problem.
For these arguments it is natural to define
the $q$-deformed covariant commutator and anticommutator,
respectively, as
\begin{eqnarray}
(q+q^{-1})(X^{+-}_{q} - X^{-+}_{q} ) & = & R_{12}A_{1}R_{21}A_{2} -
A_{2}R_{12}A_{1}R_{21} \; ,
\label{b21}
\\
(q+q^{-1})(X^{++}_{q} - X^{--}_{q} ) & = &
R_{12}A_{1}R_{21}A_{2} + A_{2}R_{12}A_{1}R^{-1}_{12} \; .
\label{b22}
\end{eqnarray}

Let us note that the
tensors (\ref{b16}) do not realize
irreps of $GL_{q}(N)$.
Indeed, contracting them over the
first or second spaces by means of $q$-traces
(\ref{b11}) we obtain
tensors transforming as in (\ref{a1}). As we have seen above,
such tensors are reduced to the 1-dimensional
and $(N^{2}-1)$-dimensional
irreps. Taking into account
these remarks we obtain finally
the following decomposition (cf. with
\cite{Smorod,Bog})
\begin{eqnarray}
X^{++}_{q} \; : \quad \frac{N^{2}(N+1)^{2}}{4} & = &
1 \oplus (N^{2} - 1) \oplus \frac{N^{2}(N+3)(N-1)}{4} \; ,
\label{b23a} \\
X^{--}_{q} \; : \quad \frac{N^{2}(N-1)^{2}}{4} & = &
1 \oplus (N^{2}-1) \oplus \frac{N^{2}(N+1)(N-3)}{4} \; ,
\label{b23b} \\
X^{\pm \mp }_{q} \; : \quad \frac{N^{2}(N^{2}-1)}{4} & = &
(N^{2}-1) \oplus \frac{(N^{2}-1)(N^{2}-4)}{4} \; .
\label{b23c}
\end{eqnarray}
We stress here that $(N^{2}-1)$- and
$\frac{N^{2}(N+1)(N-3)}{4}$-dimensional irreps appear in (\ref{b23b})
only for $N \geq 3$ and $N \geq 4$, respectively, while
$\frac{(N^{2}-1)(N^{2}-4)}{4}$-dimensional irrep appears in
(\ref{b23c}) only for $N \geq 3$.
Using the decomposition (\ref{b23a})-(\ref{b23c})
one can deduce that the direct product of two
$q$-traceless tensors can be decomposed into irreps of the following
dimensions (here $N \geq 4$):
\begin{eqnarray}
(N^{2}-1)^{\otimes 2} & = &
\left[ 1 \right] \oplus 2 \cdot \left[ N^{2}-1 \right] \oplus
\left[ \frac{(N^{2}-1)(N^{2}-4)}{4} \right]
\oplus \nonumber \\
\left [ \frac{(N^{2}-1)(N^{2}-4)}{4} \right ] ^{*}  & \oplus &
 \left[ \frac{N^{2}(N+3)(N-1)}{4} \right]
 \oplus \left[ \frac{N^{2}(N+1)(N-3)}{4} \right]  .
\label{b24}
\end{eqnarray}
In terms of the Young tableaux this formula looks like
\begin{equation}
\!\!\!\!\!\!\!\!\!\! \left(
\begin{tabular}{|c|c|}
\hline
$\!\!\!\!\! $ {\tiny 1} $\!\!\!\!\! $    & $\:$ \\
\hline
$\!\!\!\!\! \vdots \!\!\!\!\! $          &\multicolumn{1}{c}{ } \\
\cline{1-1}
$\!\!\!\!\! $ {\tiny N-1 } $\!\!\!\!\! $ & \multicolumn{1}{c}{ } \\
\cline{1-1}
\end{tabular}
\right)^{\otimes 2}
\!\! = \bullet \; \oplus \; 2 \,
\begin{tabular}{|c|c|}
\hline
$\!\!\!\!$ {\tiny 1} $\!\!\!\!$     & $\:$ \\
\hline
$\!\!\!\! \vdots \!\!\!\!$          &\multicolumn{1}{c}{ } \\
\cline{1-1}
$\!\!\!\!$ {\tiny N-1} $\!\!\!\!$   &\multicolumn{1}{c}{ } \\
\cline{1-1}
\end{tabular}
\oplus \;
\begin{tabular}{|c|c|c|}
\hline
$\!\!\!\!$ {\tiny 1} $\!\!\!\!$     & $\:$ & $\:$ \\
\hline
$\!\!\!\! \vdots \!\!\!\!$          &\multicolumn{2}{c}{ } \\
\cline{1-1}
$\!\!\!\!$ {\tiny N-2} $\!\!\!\!$   &\multicolumn{2}{c}{ } \\
\cline{1-1}
\end{tabular}
\oplus \;
\begin{tabular}{|c|c|c|}
\hline
$\!\!\!\!$ {\tiny 1} $\!\!\!\!$
& $\!\!\!\!$ {\tiny 1} $\!\!\!\!$ & $\:$ \\
\hline
$\!\!\!\!$ {\tiny 2} $\!\!\!\!$
& $\!\!\!\!$ {\tiny 2} $\!\!\!\!$ & $\:$ \\
\hline
$\!\!\!\! \vdots \!\!\!\!$
& $\!\!\!\! \vdots \!\!\!\!$  & \multicolumn{1}{c}{ } \\
\cline{1-2}
$\!\!\!\!$ {\tiny N-1} $\!\!\!\!$
& $\!\!\!\!$ {\tiny N-1} $\!\!\!\!$ &
\multicolumn{1}{c}{ } \\
\cline{1-2}
\end{tabular} \; \oplus \;
\begin{tabular}{|c|c|c|c|}
\hline $\!\!\!\!$ {\tiny 1} $\!\!\!\!$
& $\!\!\!\!$ {\tiny 1} $\!\!\!\!$ & $\:$ & $\:$ \\
\hline
$\!\!\!\! \vdots \!\!\!\!$
& $\!\!\!\! \vdots \!\!\!\!$  & \multicolumn{2}{c}{ } \\
\cline{1-2}
$\!\!\!\!$ {\tiny N-1} $\!\!\!\!$
& $\!\!\!\!$ {\tiny N-1} $\!\!\!\!$
& \multicolumn{2}{c}{ } \\
\cline{1-2}
\end{tabular}
\oplus \;
\begin{tabular}{|c|c|}
\hline
$\!\!\!\!$ {\tiny 1} $\!\!\!\!$ & $\:$ \\
\hline
$\!\!\!\!$ {\tiny 2} $\!\!\!\!$ & $\:$ \\
\hline
$\!\!\!\! \vdots \!\!\!\!$  & \multicolumn{1}{c}{ } \\
\cline{1-1}
$\!\!\!\!$ {\tiny N-2} $\!\!\!\!$ & \multicolumn{1}{c}{ } \\
\cline{1-1}
\end{tabular}
\; .
\label{b25}
\end{equation}
The dimensions of the irreps related
to the Young tableaux listed in (\ref{b25}) are
given by the Weyl formula \cite{Weyl}. Naturally, they coincide
with that expressed in Eq.(\ref{b24}).

As it will be seen in the next Section,
this information is enough to
conclude that "fermionic" (with $q$-deformed anticommutators
(\ref{b22}) in the l.h.s. of (\ref{a2})) and
"bosonic" (with $q$-deformed commutators (\ref{b21}) in the l.h.s.
of (\ref{a2})) quantum
algebras are defined uniquely up
to some inessential rescaling factors.
Moreover, we show that up to some arbitrariness
disscussed below there are no other
well defined $GL_{q}(N)$-covariant algebras  with
quadratic polynomial
structure relations (\ref{a2}).

\section{ $GL_{q}(N)$-covariant quantum algebras.}
\setcounter{equation}0

In this Section, using the R-matrix approach
\cite{FRT} we discuss the Jordan-Schwinger (J-S)
construction for covariant quantum algebras.
This is the most simple way to
reproduce explicitly quadratic polynomial
relations (\ref{a2}) for the generators of these algebras.
We start with the
formulation of the $GL_{q}(N)$-covariant
differential
calculus \cite{WesZum} on a bosonic (fermionic)
quantum hyperplane.
Commutation relations for hyperplane
coordinates and derivatives
are identical with the commutation relations for
the $GL_{q}(N)$-covariant $q$-(super)oscillators
\cite{PuszWor}--\cite{Kulish}. It is known (see
e.g. \cite{ChKL,Vokos} and  Refs. therein)
that the generators of the
quantum algebras $U_{q}(gl(N))$ can
be constructed as  bilinear combinations of
the bosonic or fermionic $q$-oscillators
(J-S construction).
In this Section,
following the idea of J-S construction we realize
the covariant quantum algebra generators
$A^{i}_{j}$ as bilinears of the $GL_{q}(N)$-covariant
$q$-oscillators.

It is known \cite{Manin,FRT} that the bosonic (fermionic)
hyperplanes with coordinates
$\{x^{i}\} =|x\rangle$
(i=1,2,\ldots,N) can be
defined by using the projectors (\ref{b10})
\begin{equation}
\label{c2a}
(R_{12} - cP_{12}) |x\rangle _{1} |x\rangle _{2}  =  0 \; ,
\end{equation}
Here $c=q$ and $c= -q^{-1}$ for bosonic
and fermionic coordinates, respectively.
Relations (\ref{c2a}) are covariant
under the left rotations of vectors $|x\rangle$
by the matrix $T^{i}_{ j}\in GL_{q}(N)$
( $|x\rangle$ is
the space of the
fundamental representation of
$GL_{q}(N)$):
\begin{equation}
x^{i} \rightarrow T^{i}_{j} x^{j} \; .
\label{c3}
\end{equation}
One can extend the
algebra (\ref{c2a}) introducing the
dual vector $\langle\partial|={\partial_{i}}$
with the transformation rule
\begin{equation}
\partial _{i} \rightarrow \partial _{j} (T^{-1})^{j}_{i} \quad , \quad
\label{c4}
\end{equation}
Then the covariant associative extension of the algebra
(\ref{c2a}) is
\begin{eqnarray}
R_{12} |x\rangle _{1} |x\rangle _{2} & = &
c |x\rangle _{2}  |x\rangle _{1} \; , \quad
\langle \partial |_{1} \langle \partial |_{2} R_{12} =
c \langle \partial |_{2} \langle \partial |_{1} \; ,
\label{c5a} \\
|x\rangle _{1} \langle \partial |_{2} & = &
\nu \delta _{12} + c\langle \partial |_{2} R_{12} |x\rangle _{1} \; .
\label{c5c}
\end{eqnarray}
Here $\delta_{12}=\delta^{i_{1}}_{j_{2}}$
is a unit matrix and $\nu$ are
arbitrary rescaling factors ($\nu = b$ for bosons and $\nu = f$ for
fermions).
Note that making the  replacements
$R_{12}\rightarrow R^{-1}_{21},
c\rightarrow c^{-1}$ in Eqs.(\ref{c5a}),(\ref{c5c})
 we obtain another
(and the last) possible covariant extension of (\ref{c2a}).
Below, we
concentrate only on the consideration of the
algebra (\ref{c5a}),(\ref{c5c})
(the other possibility can be treated analogously).

In the bosonic case ($c=q$) the
formulas (\ref{c5a}) and (\ref{c5c}) define the
covariant $q$-oscillators \cite{PuszWor} or
covariant differential calculus on the quantum hyper-plane
\cite{WesZum}. This algebra can be
interpreted also as differential calculus on
the paragrassmann hyperplane \cite{FIK} or as finite dimensional
Zamolodchikov-Faddeev algebra \cite{Kulish,Kul2}.
In the fermionic case ($c=-q^{-1}$)
the algebra (\ref{c5a}) and (\ref{c5c})
defines covariant fermionic $q$-oscillators or fermionic part of the
covariant super $q$-oscillators \cite{ChKL}.

Now, we recall that the coordinates
$\{ x^{i} \}$ and the derivatives $\{ \partial_{i} \}$
(as vector spaces) are tensors realizing  the fundamental
and contragradient representations of $GL_{q}(N)$
(see (\ref{c3}) and (\ref{c4})).
The higher order tensors
can be constructed as  direct
products of the vectors $|x \rangle $ and
$\langle \partial |$. The simplest tensor of
that kind is
\begin{equation}
A^{i}_{j} = x^{i}\partial _{j} \; .
\label{c13}
\end{equation}
The transformation rule for
this tensor coincides with (\ref{a1}) and, thus,
$A$ realizes the adjoint representation
of $GL_{q}(N)$ both for bosonic and fermionic cases.
Using formulas (\ref{c5c}) and (\ref{c13})  we obtain equation
\begin{equation}
cA_{1}R_{21}A_{2} + \nu A_{1}P_{12} =
|x\rangle _{1}|x\rangle _{2}\langle\partial |_{1}\langle\partial _{2}| \; .
\label{c14}
\end{equation}
Then, applying (\ref{c5a}) to the
right-hand side of  (\ref{c14}) we deduce the following
two relations for the operators $A^{i}_{\; j}$
\begin{eqnarray}
(R_{12}-cP_{12})(cA_{1}R_{21}A_{2} + \nu A_{1} P_{12}) & = & 0 \; ,
\label{c15a} \\
(cA_{2}R_{12}A_{1} + \nu A_{2}P_{12})(R_{21}-cP_{21}) & = & 0 \; .
\label{c15b}
\end{eqnarray}
Difference between (\ref{c15a}) and (\ref{c15b})
gives the $q$-deformed commutation relations
(cf. with (\ref{b21}))
\begin{equation}
R_{12}A_{1}R_{21}A_{2} - A_{2}R_{12}A_{1}R_{21} =
\mu (P_{12}A_{1}R_{21} - R_{12}A_{1}P_{12}) \; ,  \;\;
\mu = \frac{\nu}{c} \; .
\label{c16}
\end{equation}
By construction, these relations are covariant under
the adjoint $GL_{q}(N)$-coaction (\ref{a1}).
Note that the algebra (\ref{c16}) is the same for bosonic and
fermionic $q$-oscillators (up to some trivial rescaling of
the generators $A^{i}_{j}$).
In the classical limit $q=1$, Eqs.(\ref{c16})
coinside with the
usual commutation relations for the $gl(N)$-algebra.
We call the algebra
with the structure relations (\ref{c16})
as "bosonic" $GL_{q}(N)$-covariant
quantum algebra. One can check that this algebra is associative.
The invariant central elements (Casimir
operators) for the algebra (\ref{c16})
are represented in the form (\ref{b14}). The identities
$[C_{n},A^{i}_{ j}]=0$ can be
obtained by using the Hecke relation (\ref{b9a}), the property of the
$q$-trace (\ref{b13}) and
the fact that the matrix
$Tr_{2}(D_{2}P_{12}R_{12})$ is
proportional to the unit matrix in the first space.

The $q$-deformed commutation relations
(\ref{c16}) can be rewritten in the form
\begin{equation}
\label{is}
R_{12}\tilde{A}_{1}R_{21}\tilde{A}_{2}
- \tilde{A}_{2}R_{12}\tilde{A}_{1}R_{21} = \kappa
(P_{12}\tilde{A}_{1}R_{21} - R_{12}\tilde{A}_{1}P_{12}) \; ,
 \;\; [ H, \; \tilde{A}^{i}_{j} ] =0 \; ,
\end{equation}
$$
\kappa =
\mu + \frac{(q-q^{-1})^{2}}{(q^{N}-q^{-N})}H \; .
$$
Here $\tilde{A}^{i}_{j}$ are the $q$-traceless generators
(see (\ref{b11})) and
$H=q^{-N-1}Tr_{q}(A)$.
Thus, the algebra (\ref{c16}) is the direct sum of the trivial algebra
generated by the central element $H$ and the algebra generated by
the operators $\tilde{A}^{i}_{j}$. As we will see below,
the operators $\tilde{A}^{i}_{j}$ and $A^{i}_{j}$
can be interpreted as
invariant vector fields on the $SL_{q}(N)$ and
$GL_{q}(N)$, respectively.
Finally, we rewrite the relations (\ref{c16})
in the form
\begin{equation}
R_{12}Y_{1}R_{21}Y_{2} - Y_{2}R_{12}Y_{1}R_{21} = 0 \; ,
\label{c17}
\end{equation}
where
$
A^{i}_{j} = \frac{- \mu}{(q-q^{-1})}\delta^{i}_{j} + Y^{i}_{j} \; .
$
Eq.(\ref{c17}) is well known
as reflection equation \cite{Kul2} or as relations for the operator
$Y=(L^{-})^{-1}L^{+}$,
where the elements of triangular matrices $L^{\pm}$ are
defined by the generators of
the Borel subalgebras  of $U_{q}(gl(N))$ (see \cite{FRT}).
In Refs. \cite{Woron2}-\cite{Faddeev} the operator Y
is interpreted as differential operators (vector fields) of the
bicovariant differential calculus on $GL_{q}(N)$.
The algebra (\ref{c17}) is known also as the braided algebra
\cite{Mad}. We present here also the commutation relations
of $Y$ with
$\langle\partial |$ and $|x\rangle$
$$
|x\rangle_{1} Y_{2} = R_{21}Y_{2}R_{12}|x\rangle_{1} \; , \;\;
Y_{2}\langle\partial |_{1} = \langle\partial |_{1}R_{21}Y_{2}R_{12} \; .
$$

We have considered only part of the relations (\ref{c15a}) and (\ref{c15b}),
namely the relations (\ref{c16}). Now we
proceed to the discussion of the rest of Eqs. (\ref{c15a}),
(\ref{c15b}). First of all we rewrite them in the equivalent
form
\begin{eqnarray}
(R_{12} - c P_{12})(cA_{1}R_{21}A_{2} + \nu A_{1}P_{12})
(R_{12} - c P_{12}) & = & 0 \; ,
\label{c20a} \\
(R_{12} \mp c^{\pm 1} P_{12})(cA_{1}R_{21}A_{2} + \nu A_{1}P_{12})
(R_{12} \pm c^{\mp 1} P_{12}) & = & 0 \; .
\label{c20b}
\end{eqnarray}
The pair of Eqs. (\ref{c20b}) are equivalent to the
commutation relations (\ref{c16}) for the "bosonic" $GL_{q}(N)$-covariant
quantum algebra.
Indeed, acting on (\ref{c16}) by the projectors
$(R_{12} \pm c^{\mp 1}P_{12})$ from the
left we obtain
(\ref{c20b}). On the other hand,
difference between two of Eqs.(\ref{c20b}) gives
(\ref{c16}).
The remaining relation (\ref{c20a}) takes the different forms
for the bosonic and fermionic oscillators. For
the bosonic case we obtain
\begin{equation}
\label{c21a}
(R_{12} - q P_{12})(A_{1}R_{21}A_{2} + bq^{-1} A_{1}P_{12})
(R_{12} - q P_{12})  = 0 \; ,
\end{equation}
while for the fermionic case we have
\begin{equation}
\label{c21b}
(R_{12} + q^{-1} P_{12})(A_{1}R_{21}A_{2} - fq A_{1}P_{12})
(R_{12} + q^{-1} P_{12})  =  0 \; .
\end{equation}
The bilinear parts of
Eqs.(\ref{c21a}),(\ref{c21b}) coinside with
$X_{q}^{--}P_{12}$ and $X_{q}^{++}P_{12}$, respectively
(see (\ref{b16})) and,
hence, combining these equations together
we shall obtain
$GL_{q}(N)$-covariant relations
with the $q$-deformed anticommutator (\ref{b22}).
Indeed, subtracting (\ref{c21a}) from (\ref{c21b}) we deduce
$$
R_{12}A_{1}R_{21}A_{2} + A_{2}R_{12}A_{1}R_{12}^{-1} =
$$
\begin{equation}
\label{c21f}
P_{12} \left( q^{-1}b P^{-}_{12} +
qf P^{+}_{12}
\right) (A_{1}R_{21} + R_{12}^{-1}A_{2}) =
\nu \left( R_{12}A_{1}R_{21} + A_{2} \right) \; .
\end{equation}
We interpret (\ref{c21f}) as structure relations for
"fermionic" $GL_{q}(N)$-covariant algebra and we are obliged
to put $b=f=\nu$ in order to have the associative algebra.
The contraction
$b=0, \; f=0$ of the algebra (\ref{c21f}) leads to
the relations
\begin{equation}
\label{c21g}
R_{12}A_{1}R_{21}A_{2} + A_{2}R_{12}A_{1}R_{12}^{-1} = 0 \; ,
\end{equation}
which, as we will see below,
are the $q$-deformed anticommutation relations for the
Cartan's 1-forms on the $GL_{q}(N)$. Note that the relations
(\ref{c21f}) can be rewritten in the form
\begin{equation}
R_{12}W_{1}R_{21}W_{2} + W_{2}R_{12}W_{1}R^{-1}_{12} =
\frac{\nu^{2}}{2} (R_{12}R_{21} +1) \; ,
\end{equation}
where $A^{i}_{j}=\frac{\nu}{2}\delta^{i}_{j} + W^{i}_{j}$.

The logic of J-S construction
allows us in principle to change the
$q$-deformed commutation relations
(\ref{c16}) by mixing them with
the additional relations
(\ref{c20a}).
The existense of these additional relations has been
pointed out in Ref.\cite{Vokos} where J-S construction has
been considered in the noncovariant way.
But it is natural to demand the covariance of
$q$-commutation relations under the transformation
(\ref{a1}). This remark
and the requirements discussed in the
previous Sections impose very strong
restrictions on the possible form of $q$-commutation
relations. It seems that the only reasonable choices here are
those of (\ref{c16}) and (\ref{c21f}).
However, there is
remaining abitrariness which
now we have to discuss.

Covariant relations
(\ref{c16}) and (\ref{c21f}) define the covariant "bosonic"
and "fermionic" algebras which are "good" in the sense that they allow
to reorder  any monomial $A^{i}_{j} \dots A^{k}_{l}$.
But these relations are not the only possible
covariant relations of the kind (\ref{a2}).
It is clear that (\ref{c16}) and (\ref{c21f}) are  linear
combinations of the "irreducible" sets of covariant relations
(ISCR) which correspond
to the irreps presented in
(\ref{b23a})-(\ref{b25}). Note that among these ISCR there are
several independent "adjoint" ISCR, namely
a couple of trivial "adjoint" ISCR
($[\tilde{A},Tr_{q}(A)]_{\pm}$)
and a couple, for $N \geq 3$ (or one, for $N=2$),
of nontrivial ones (see (\ref{b25})).
Some linear combinations of
these "adjoint" ISCR are included in both
the "bosonic" and "fermionic"
covariant algebras. Their presence is evident due to
the existence of linear terms in the formulas
(\ref{c16}) and (\ref{c21f}).
Leaving aside here the problem of the associativity
one can use the different combinations of
the "adjoint" ISCR instead of original ones in the covariant relations
(\ref{c16}) and (\ref{c21f}).
The only restriction is that
these combinations must contain both
the trivial and nontrivial "adjoint" ISCR
(to solve the problem of ordering).
However,
it is rather difficult to write
the new algebras in the compact form.
So, the covariant algebras (\ref{c16}) and (\ref{c21f}) look
preferable.

To conclude this Section, we illustrate our results by considering,
in detail, the special case
of $N=2$. For this we introduce
the new notation
\begin{equation}
A^{i}_{j}\; = \;
\left(
\begin{array}{cc}
A^{1}_{1} & A^{1}_{2} \\
A^{2}_{1} & A^{2}_{2}
\end{array}
\right) \; = \;
\left(
\begin{array}{cc}
\frac{H+qA_{0}}{q+q^{-1}} & A_{+} \\
A_{-} & \frac{H-q^{-1}A_{0}}{q+q^{-1}}
\end{array}
\right)
\label{c26}
\end{equation}
where $H=q^{-3}Tr_{q}A=(q^{-1}A^{1}_{1}+qA^{2}_{2})$
and $A_{0}=A^{1}_{1}-A^{2}_{2}$.

The $GL_{q}(2)$-covariant
"bosonic" quantum algebra (\ref{c16}) is rewritten as
(we change the notation $A$ to $E$
bearing in mind the interpretation of the
matrix elements (\ref{c26}) as invariant
vector fields on $GL_{q}(2)$)
\begin{eqnarray}
[E_{-}, E_{+}] & = &
\frac{q^{2}-1}{q^{2}+1} E^{\;2}_{0} + \frac{\kappa}{q}E_{0} \; ,
\label{c27a} \\
\mbox{[} E_{\pm}, E_{0} \mbox{]} _{(q^{\mp 1},q^{\pm 1})} \; & \equiv & \;
q^{\mp 1}E_{\pm}E_{0} - q^{\pm 1}E_{0}E_{\pm}
\; = \; \pm (q+q^{-1}) \frac{\kappa}{q}E_{\pm} \; ,
\label{c27b} \\
\mbox{[}H,E_{\pm} \mbox{]} & = & \mbox{[}H,E_{0}\mbox{]} \; = \; 0 \; ,
\label{c27c}
\end{eqnarray}
where $\kappa$ is defined in (\ref{is}) for $N=2$.
Performing the transformations (\ref{a1}) for the generators (\ref{c26})
we may directly convince ourselves
that the relations (\ref{c27a})-(\ref{c27b})
define the covariant algebra.
The central element $H=q^{-3}Tr_{q}A$ of the algebra
(\ref{c27a})-(\ref{c27c})
can be removed by the following rescaling
$ E_{\pm ,0}  = (1+q^{-2})\kappa \hat{E}_{\pm ,0}$
and finally we obtain
\begin{equation}
(q+q^{-1})\mbox{[} \hat{E}_{-}, \hat{E}_{+} \mbox{]}  -
(q-q^{-1})\hat{E}^{\;2}_{0} = \hat{E}_{0} \quad , \quad
\mbox{[} \hat{E}_{\pm} , \hat{E}_{0} \mbox{]}
_{(q^{\mp 1}, q^{\pm 1})}  =
\pm \hat{E}_{\pm} \; .
\label{c28}
\end{equation}
These relations
really correspond to the adjoint irrep
(
\begin{tabular}{|c|c|}
\hline
$\:$ & $\:$ \\
\hline
\end{tabular}
) of $SL_{q}(2) \subset GL_{q}(2)$.
As a covariant object, the algebras
(\ref{c27a})-(\ref{c27c})
and (\ref{c28}) have been considered in \cite{Isaev}.
Note that up to some trivial rescalings
the commutation relations (\ref{c28})
coincide with those for Witten's deformation
of the algebra $sl(2)$
(see Eqs.(5.2) of \cite{Witten}).

The defining relations for the $GL_{q}(2)$-covariant "fermionic"
quantum algebra looks like (we change
the notation $A^{i}_{j}$ to $\Omega^{i}_{j}$
bearing in mind the interpretation of this matrix elements
as Cartan's 1-forms on $GL_{q}(2)$):
\begin{equation}
\label{c29a}
\begin{tabular}{|c|c|c|c|}
\hline $\:$ & $\:$ & $\:$ & $\:$ \\
\hline
\end{tabular} \; : \quad
\left\{
\begin{array}{c}
q^{2}\Omega_{+} \Omega_{-} + q^{-2}\Omega_{-} \Omega_{+} -
\Omega^{\;2}_{0}  =  0 \; ,
\\
q^{\mp1}\Omega_{0} \Omega_{\pm} + q^{\pm1}\Omega_{\pm} \Omega_{0}
 =  0 \quad , \quad
\Omega^{\;2}_{\pm}  = 0 \; ;
\end{array}
\right.
\end{equation}
\begin{equation}
\label{c29b}
\!\!\!\!\!\! \begin{array}{l}
r \; \begin{tabular}{|c|c|}
\hline
$\:$ & $\:$ \\
\hline
\end{tabular} \; + \\
 + \left\{
H,\Omega
\right\}
\end{array}
 \: : \:
\left\{
\begin{array}{l}
r \left(
(q+\frac{1}{q}) [\Omega_{-},\Omega_{+}] -
(q-\frac{1}{q})\Omega^{2}_{0}
\right) +
\left\{ H, \Omega_{0} \right\}  = \lambda \Omega_{0}  , \\
r \left[ \Omega_{\pm}, \Omega_{0} \right]_{(q^{\mp 1},q^{\pm 1})}
\pm
\left\{ H , \Omega_{\pm} \right\}
 = \pm \lambda \Omega_{\pm}  ;
\end{array}
\right.
\end{equation}
\begin{equation}
\label{c29c}
the \; scalar \; ISCR \; : \quad
\left\{
\begin{array}{c}
-\frac{1}{q} Tr_{q}(\Omega^{2}) + q H^{2} = \nu H \; , \\
Tr_{q}(\Omega^{2}) + H^{2} = q(q^{2} + 1 + q^{-2})\nu H \; .
\end{array}
\right.
\end{equation}
Here $\lambda = (q+q^{-1})\nu$ , $r=(1-q^{2})/(q^{2}+q^{-2})$ and
\begin{equation}
\label{c30}
\frac{1}{q^{3}}Tr_{q}(\Omega^{2}) \; = \;
\left( q^{-1}\Omega_{+}\Omega_{-} + q\Omega_{-}\Omega_{+} \right)
+ \frac{\Omega^{\;2}_{0} + H^{2}}{q+q^{-1}} \; ,
\end{equation}
In the limit $\nu=0$,
Eqs. (\ref{c29a})-(\ref{c29c}) become the commutation
relations for the Cartan's 1-forms on $GL_{q}(2)$.
These relations in another form have been
presented in Ref. \cite{Zumino}.
Note that just the presence of the ISCR
\begin{tabular}{|c|c|}
\hline
$\:$ & $\:$ \\
\hline
\end{tabular}
in these relations prevents us (for $q \neq 1$) to remove
$H$ and pass over to the Cartan's 1-forms on $SL_{q}(2)$
\footnote{This feature was pointed out in Ref.\cite{Zumino}
for $SL_{q}(2)$
and in Ref.\cite{Woron2} in general.}.
We believe that the right way to obtain the commutation relations
for Cartan's 1-forms on $SL_{q}(2)$ is simply to ignore the
Eqs. (\ref{c29b}) and use only the Eqs. (\ref{c29a})
and (\ref{c29c}) for $H=0$.
These relations define the associative covariant algebra and
have the correct classical limit.

Finally, one can check directly that the quadratic Casimir operators
for the algebras (\ref{c27a}),(\ref{c27b}) and (\ref{c29a})-(\ref{c29c})
are related to the invariant $C_{2}$ (see (\ref{b14}) and (\ref{c30})).

\section{Conclusion}
\setcounter{equation}0

To conclude, we present here
an explicite construction for the invariant
vector fields and 1-forms on $GL_{q}(N)$ and thus illustrate
the connection between $GL_{q}(N)$-covariant
quantum algebras and the covariant differential calculus
on $GL_{q}(N)$.
Let us introduce the quantum group derivatives
$\partial^{i}_{j} = \partial / \partial T^{j}_{i}$
and differentials $d(T^{i}_{j})$ to extend $GL_{q}(N)$
in the following way
\begin{eqnarray}
R_{12}T_{1}T_{2} \; = \; T_{2}T_{1}R_{12} & , &
R_{12}\partial_{2}\partial_{1} \; = \; \partial_{1}\partial_{2}R_{12} \; ,
\nonumber \\
\partial_{2} R_{12}T_{1} \; = \;
\nu P_{12} + T_{1}R^{\;-1}_{21}\partial_{2} & , &
R_{21}^{\;-1}T_{1}d(T_{2}) \; = \;
d(T_{2})T_{1}R_{12} \; ,
\nonumber \\
R_{21}^{\;-1}d(T_{1})d(T_{2}) & = &
-d(T_{2})d(T_{1})R_{12} \; .
\label{d1}
\end{eqnarray}
This algebra is covariant under
the left(right) $GL_{q}(N)$-coaction on the operators
$T$, $\partial$ and $d(T)$ which can be considered as
bicomodules of $GL_{q}(N)$:
$T  \rightarrow  \tilde{T}T T'$,
$\partial  \rightarrow
 T'^{-1}
\partial \tilde{T}^{-1} $,
$
d(T)  \rightarrow  \tilde{T}d(T) T'$,
where $\tilde{T}^{i}_{j}, \; T'^{i}_{j}$ are generators
of various examples of $GL_{q}(N)$.
Using the relations (\ref{d1}) one can directly
check that operators $E=T\partial$ satisfy "bosonic" commutation
relations (\ref{c17}) and operators $\Omega=d(T)T^{-1}$ satisfy
contracted "fermionic" anticommutation relations (\ref{c21g})
\footnote{Note that the relations (\ref{c21g})
are also covariant under
the "gauge" coaction $\Omega \rightarrow T\Omega T^{-1} +d(T)T^{-1}$.}.
Thus, we
relate the $GL_{q}(N)$-covariant quantum algebras introduced in
the previous Section with  bicovariant differential calculus on
$GL_{q}(N)$.
\section*{Acknowledgments}
We thank R.Kashaev, R.P.Malik and V.Tolstoy for
discussions and comments.
One of the
authors (API) had extremely
useful discussions with
Prof. Ya.A.Smorodinsky when this paper was in progress.
We are very sad to know his sudden demise.

\end{document}